# Modern Comprehensive study of the W UMa System TY Boo

M. M. Elkhateeb[1,2], M. I. Nouh[1,2] and A. S. Saad[1,3]

[1]Astronomy Department, National Research Institute of Astronomy and Geophysics, 11421 Helwan, Cairo, Egypt

E-mail: abdo_nouh@hotmail.com, Fax: +202 2554 8020

[2]Physics department, College of Science, Northern Border University, 1321 Arar, Saudi Arabia

[3]Department of Mathematics, Preparatory Year, Qassim University, Buraidah, Saudi Arabia.

**Abstract:** New CCD three colors light curves of TY Boo were carried out in five nights from February to May 2006 in BVR bandpass using a 50-cm F/8.4 Ritchey–Chretien telescope (Ba50) of the Baja Astronomical Observatory (Hungary), and 512 × 512 Apogee AP-7 CCD camera. A photometric solution of these light curves was obtained by means of Willson-Devinney code. The results showed that the low massive component is hotter than the more massive and cool one, and the temperature difference between the components is $\Delta T \sim 249$ K. Long term investigation for the period behavior of the system based on all available data shows two stages of increase and similar of decrease, which look like as a periodic behavior. A new light elements yields a new period (p= $0.^d3171506$) and shows a period decrease with the rate $dP/dE = 5.858 \times 10^{-12}$ day/cycle or $6.742 \times 10^{-9}$ day/year or 0.1 second/century. The evolutionary status of the system is discussed.

1. **Introduction**

The eclipsing binary system TY Boo was discovered as a variable star and classified as a W UMa type by Guthnick and Prager (1926), with a period of $0.^d31730$. A cyclic period variation of about 400 orbital revolutions (127 days) was found by Szafraniec (1953). Carr (1972) suggested that the system was an A-type system consisting of two main sequence (G3 and G7) components. The published data by Carr (1972) was re-analyzed by Niarchos (1978) using frequency domain techniques. The results suggested that the system was a W UMa system with a mass ratio of 0.22. A new BV light curves was published by Samec&Bookmyer (1987), they concluded that the system becomes redder since Carr's observations, and the depths of the eclipse curves didn't show any apparent changes. Their period study shows no indication of the cyclic period variation which suggested by Szafraniec (1953).

The first Spectroscopic observations for the system TY Boo was carried out by Rainger et al. (1990). They used their spectroscopic observations together with B light curves published by Samec&Bookmyer (1987) to yield a combined orbital solution, which gave the masses and absolute dimensions for the components. Their results show that the system is a normal W type contact binary, with a main sequence primary star and a secondary component larger than expected for its ZAMS mass by ~ 1.4. They confirmed the orbital period change of the system, and calculated the first radial velocities for the system.

Photometric and spectroscopic observations were carried out for the system TY Boo by Milone et al. (1991) in three observing seasons. They calculated the mass ratio of the system (q = $M_h/M_c$ = 0.465) which is consistent with the value derived using light curve analysis (q = 0.481). Christopoulou et al. (2012) observed the system in four color bandpass and displayed a long term light curve solution.

In this work we present a new CCD light curves in BVR band for the system TY Boo which analyzed using an advanced version of the Wilson and Devinney synthetic light curve code. Long term stability for both period and observed light curves was investigated and a possible connection between them was searched.

## 2. Observations

The present CCD Observations of TY Boo were carried out on five nights from February to May 2006 in BVR bandpass using a 50-cm F/8.4 Ritchey–Chretien telescope (Ba50) of the Baja Astronomical Observatory (Hungary), and 512 × 512 Apogee AP-7 CCD camera.

The observed frames were processed by the photometry software AIP4WIN (Berry and Buruell 2000) which based on aperture photometry, including bias and dark subtraction and flat field correction. Star GSC 02568-00997 (V=11.59 mag, B-V = 0.37) was used as a comparison star, while GSC 02568-00991 (V=11.67 mag) used as a check. The original data were listed in Table 1. A total of 962 individual observations were obtained in BVR bandpass (240 in B, 293 in V and 429 in R). The BVR light curves displayed in Figure 1 show the magnitude difference (the variable minus the comparison star) versus phase in BVR bands. The orbital phases were computed according to the following ephemeris by Kreiner et al. (2001):

$$\text{Min I} = 2447612.6035 + 0.3171490 \qquad (1)$$

Figure 1, indicates that the light curve variation of the system TY Boo is typical of W UMa type and the data of all days joined smoothly. A new 19 times of minima were derived (12 primary and 7 secondary)

were estimated by means of the Minima V2.3 package (Nelson, 2006) based on the Kwee&Van Worden (1956) fitting method. The new minima appear in Table 2 and together in Table 3 with all photometric and CCD timings which have been published.

Table 1: The magnitude difference in BVR band of TY Boo together with the heliocentric Julian dates and phases.

| B-band observations of TY Boo | | | | V-band observations of TY Boo | | | | R-band observations of TY Boo | | | |
|---|---|---|---|---|---|---|---|---|---|---|---|
| JD | Phase | ΔB | Error | JD | Phase | ΔV | Error | JD | Phase | ΔR | Error |
| 2453796.4316 | 0.0132 | -0.621 | 0.025 | 2453794.4926 | 0.7794 | -0.863 | 0.035 | 2453794.5147 | 0.9810 | 0.087 | 0.004 |
| 2453796.4342 | 0.0215 | -0.640 | 0.026 | 2453794.4942 | 0.7845 | -0.872 | 0.036 | 2453794.5163 | 0.9741 | 0.114 | 0.005 |
| 2453796.4369 | 0.0299 | -0.649 | 0.027 | 2453794.4974 | 0.7946 | -0.860 | 0.035 | 2453794.5179 | 0.9791 | 0.128 | 0.005 |
| 2453796.4395 | 0.0383 | -0.712 | 0.029 | 2453794.4990 | 0.7997 | -0.863 | 0.035 | 2453794.5200 | 0.9857 | 0.158 | 0.007 |
| 2453796.4422 | 0.0466 | -0.790 | 0.032 | 2453794.5006 | 0.8048 | -0.860 | 0.035 | 2453794.5216 | 0.9907 | 0.146 | 0.006 |
| 2453796.4448 | 0.0549 | -0.831 | 0.034 | 2453794.5023 | 0.8099 | -0.835 | 0.034 | 2453794.5261 | 0.0051 | 0.158 | 0.007 |
| 2453796.4475 | 0.0633 | -0.903 | 0.037 | 2453794.5039 | 0.8149 | -0.831 | 0.034 | 2453794.5309 | 0.0203 | 0.140 | 0.006 |
| 2453796.4501 | 0.0717 | -0.955 | 0.039 | 2453794.5055 | 0.8200 | -0.850 | 0.035 | 2453794.5325 | 0.0253 | 0.120 | 0.005 |
| 2453796.4528 | 0.0800 | -0.968 | 0.040 | 2453794.5071 | 0.8251 | -0.828 | 0.034 | 2453794.5341 | 0.0304 | 0.088 | 0.004 |
| 2453796.4554 | 0.0883 | -1.056 | 0.043 | 2453794.5087 | 0.8302 | -0.821 | 0.034 | 2453794.5358 | 0.0355 | 0.073 | 0.003 |
| 2453796.4581 | 0.0968 | -1.040 | 0.043 | 2453794.5103 | 0.8352 | -0.810 | 0.033 | 2453794.5374 | 0.0406 | 0.052 | 0.002 |
| 2453796.4607 | 0.1051 | -1.066 | 0.044 | 2453794.5119 | 0.8403 | -0.831 | 0.034 | 2453794.5390 | 0.0456 | 0.017 | 0.001 |
| 2453796.4634 | 0.1134 | -1.097 | 0.045 | 2453794.5135 | 0.8453 | -0.788 | 0.032 | 2453794.5406 | 0.0507 | -0.016 | 0.001 |
| 2453796.4660 | 0.1218 | -1.153 | 0.047 | 2453794.5151 | 0.8504 | -0.814 | 0.033 | 2453794.5422 | 0.0558 | -0.038 | 0.002 |
| 2453796.4687 | 0.1302 | -1.170 | 0.048 | 2453794.5167 | 0.8555 | -0.798 | 0.033 | 2453794.5452 | 0.0654 | -0.096 | 0.004 |
| 2453796.4713 | 0.1385 | -1.147 | 0.047 | 2453794.5183 | 0.8606 | -0.781 | 0.032 | 2453794.5469 | 0.0705 | -0.122 | 0.005 |
| 2453796.4740 | 0.1469 | -1.198 | 0.049 | 2453794.5199 | 0.8656 | -0.787 | 0.032 | 2453794.5485 | 0.0756 | -0.159 | 0.007 |
| 2453796.4766 | 0.1552 | -1.201 | 0.049 | 2453794.5216 | 0.8707 | -0.783 | 0.032 | 2453794.5501 | 0.0807 | -0.186 | 0.008 |
| 2453796.4793 | 0.1635 | -1.208 | 0.049 | 2453794.5232 | 0.8758 | -0.762 | 0.031 | 2453794.5517 | 0.0857 | -0.200 | 0.008 |
| 2453796.4819 | 0.1719 | -1.283 | 0.052 | 2453794.5248 | 0.8808 | -0.763 | 0.031 | 2453794.5533 | 0.0908 | -0.221 | 0.009 |
| 2453796.4846 | 0.1803 | -1.263 | 0.052 | 2453794.5264 | 0.8859 | -0.733 | 0.03 | 2453794.5549 | 0.0958 | -0.237 | 0.010 |
| 2453796.4872 | 0.1886 | -1.266 | 0.052 | 2453794.5280 | 0.8910 | -0.714 | 0.029 | 2453794.5774 | 0.1668 | -0.399 | 0.016 |
| 2453796.4899 | 0.1970 | -1.266 | 0.052 | 2453794.5296 | 0.8961 | -0.723 | 0.03 | 2453794.5790 | 0.1719 | -0.414 | 0.017 |
| 2453796.4925 | 0.2054 | -1.290 | 0.053 | 2453794.5312 | 0.9012 | -0.702 | 0.029 | 2453794.5822 | 0.1820 | -0.419 | 0.017 |
| 2453796.4952 | 0.2137 | -1.333 | 0.054 | 2453794.5328 | 0.9062 | -0.686 | 0.028 | 2453794.5838 | 0.1871 | -0.436 | 0.018 |
| 2453796.4978 | 0.2220 | -1.344 | 0.055 | 2453794.5344 | 0.9113 | -0.665 | 0.027 | 2453794.5855 | 0.1922 | -0.436 | 0.018 |
| 2453796.5005 | 0.2304 | -1.301 | 0.053 | 2453794.5360 | 0.9164 | -0.634 | 0.026 | 2453794.5871 | 0.1973 | -0.451 | 0.018 |
| 2453796.5031 | 0.2388 | -1.332 | 0.054 | 2453794.5377 | 0.9215 | -0.625 | 0.026 | 2453794.5887 | 0.2023 | -0.453 | 0.019 |
| 2453796.5058 | 0.2472 | -1.332 | 0.054 | 2453794.5393 | 0.9265 | -0.592 | 0.024 | 2453794.5903 | 0.2074 | -0.461 | 0.019 |
| 2453796.5084 | 0.2555 | -1.330 | 0.054 | 2453794.5409 | 0.9315 | -0.582 | 0.024 | 2453794.5919 | 0.2125 | -0.47 | 0.019 |
| 2453796.5111 | 0.2639 | -1.350 | 0.055 | 2453794.5425 | 0.9366 | -0.542 | 0.022 | 2453794.5935 | 0.2176 | -0.463 | 0.019 |
| 2453796.5137 | 0.2722 | -1.336 | 0.055 | 2453794.5427 | 0.0625 | -0.439 | 0.018 | 2453794.5951 | 0.2227 | -0.477 | 0.020 |
| 2453796.5164 | 0.2806 | -1.326 | 0.054 | 2453794.5458 | 0.0721 | -0.495 | 0.02 | 2453794.6016 | 0.2430 | -0.474 | 0.019 |
| 2453796.5293 | 0.3213 | -1.288 | 0.053 | 2453794.5474 | 0.0772 | -0.521 | 0.021 | 2453794.6031 | 0.2480 | -0.48 | 0.020 |
| 2453796.5319 | 0.3297 | -1.289 | 0.053 | 2453794.5490 | 0.0822 | -0.543 | 0.022 | 2453794.6047 | 0.2530 | -0.480 | 0.020 |
| 2453796.5346 | 0.3381 | -1.283 | 0.052 | 2453794.5506 | 0.0873 | -0.578 | 0.024 | 2453794.6063 | 0.2581 | -0.480 | 0.020 |
| 2453796.5373 | 0.3464 | -1.266 | 0.052 | 2453794.5522 | 0.0924 | -0.594 | 0.024 | 2453794.6080 | 0.2632 | -0.479 | 0.020 |
| 2453796.5425 | 0.3630 | -1.250 | 0.051 | 2453794.5538 | 0.0974 | -0.625 | 0.026 | 2453794.6096 | 0.2682 | -0.473 | 0.019 |
| 2453796.5452 | 0.3714 | -1.237 | 0.051 | 2453794.5554 | 0.1025 | -0.629 | 0.026 | 2453794.6128 | 0.2784 | -0.477 | 0.020 |
| 2453796.5478 | 0.3798 | -1.236 | 0.051 | 2453794.5570 | 0.1076 | -0.649 | 0.027 | 2453794.6144 | 0.2835 | -0.469 | 0.019 |
| 2453796.5505 | 0.3881 | -1.198 | 0.049 | 2453794.5586 | 0.1126 | -0.666 | 0.027 | 2453794.6176 | 0.2937 | -0.472 | 0.019 |
| 2453796.5531 | 0.3965 | -1.183 | 0.048 | 2453794.5602 | 0.1177 | -0.661 | 0.027 | 2453794.6209 | 0.3039 | -0.458 | 0.019 |
| 2453796.5558 | 0.4048 | -1.177 | 0.048 | 2453794.5605 | 0.9937 | -0.238 | 0.01 | 2453794.6225 | 0.3089 | -0.467 | 0.019 |
| 2453796.5584 | 0.4132 | -1.139 | 0.047 | 2453794.5619 | 0.1228 | -0.707 | 0.029 | 2453794.6241 | 0.3140 | -0.448 | 0.018 |
| 2453796.5611 | 0.4215 | -1.071 | 0.044 | 2453794.5622 | 0.9988 | -0.229 | 0.009 | 2453796.4325 | 0.0160 | 0.150 | 0.006 |
| 2453796.5637 | 0.4299 | -1.021 | 0.042 | 2453794.5635 | 0.1279 | -0.713 | 0.029 | 2453796.4351 | 0.0243 | 0.133 | 0.005 |
| 2453796.5664 | 0.4382 | -1.007 | 0.041 | 2453794.5651 | 0.1329 | -0.738 | 0.03 | 2453796.4378 | 0.0327 | 0.098 | 0.004 |
| 2453796.5690 | 0.4466 | -0.968 | 0.040 | 2453794.5667 | 0.1380 | -0.759 | 0.031 | 2453796.4404 | 0.0410 | 0.049 | 0.002 |
| 2453796.5717 | 0.4549 | -0.930 | 0.038 | 2453794.5683 | 0.1430 | -0.765 | 0.031 | 2453796.4431 | 0.0494 | 0.005 | 0.000 |
| 2453796.5743 | 0.4632 | -0.878 | 0.036 | 2453794.5699 | 0.1481 | -0.783 | 0.032 | 2453796.4457 | 0.0577 | -0.046 | 0.002 |
| 2453796.5770 | 0.4716 | -0.850 | 0.035 | 2453794.5715 | 0.1532 | -0.762 | 0.031 | 2453796.4484 | 0.0661 | -0.090 | 0.004 |
| 2453796.5796 | 0.4800 | -0.807 | 0.033 | 2453794.5715 | 0.0283 | -0.224 | 0.009 | 2453796.4510 | 0.0744 | -0.132 | 0.005 |
| 2453796.5823 | 0.4883 | -0.775 | 0.032 | 2453794.5731 | 0.1582 | -0.782 | 0.032 | 2453796.4537 | 0.0828 | -0.176 | 0.007 |
| 2453796.5849 | 0.4966 | -0.775 | 0.032 | 2453794.5731 | 0.0334 | -0.250 | 0.01 | 2453796.4563 | 0.0911 | -0.209 | 0.009 |
| 2453796.5876 | 0.5050 | -0.776 | 0.032 | 2453794.5747 | 0.1633 | -0.777 | 0.032 | 2453796.4590 | 0.0995 | -0.240 | 0.010 |
| 2453796.5902 | 0.5134 | -0.814 | 0.033 | 2453794.5748 | 0.0384 | -0.268 | 0.011 | 2453796.4616 | 0.1079 | -0.267 | 0.011 |
| 2453796.5929 | 0.5217 | -0.816 | 0.033 | 2453794.5763 | 0.1685 | -0.788 | 0.032 | 2453796.4643 | 0.1162 | -0.284 | 0.012 |
| 2453796.5955 | 0.5300 | -0.837 | 0.034 | 2453794.5764 | 0.0435 | -0.287 | 0.012 | 2453796.4669 | 0.1246 | -0.303 | 0.012 |
| 2453796.5982 | 0.5384 | -0.895 | 0.037 | 2453794.5780 | 0.1735 | -0.794 | 0.032 | 2453796.4696 | 0.1330 | -0.319 | 0.013 |
| 2453796.6008 | 0.5468 | -0.931 | 0.038 | 2453794.5780 | 0.0486 | -0.336 | 0.014 | 2453796.4722 | 0.1413 | -0.342 | 0.014 |
| 2453796.6035 | 0.5551 | -0.974 | 0.040 | 2453794.5796 | 0.1786 | -0.826 | 0.034 | 2453796.4748 | 0.1496 | -0.358 | 0.015 |
| 2453796.6061 | 0.5635 | -1.005 | 0.041 | 2453794.5796 | 0.0537 | -0.325 | 0.013 | 2453796.4775 | 0.1580 | -0.367 | 0.015 |
| 2453796.6088 | 0.5719 | -1.032 | 0.042 | 2453794.5812 | 0.1837 | -0.791 | 0.032 | 2453796.4802 | 0.1663 | -0.379 | 0.016 |
| 2453796.6114 | 0.5802 | -1.092 | 0.045 | 2453794.5812 | 0.0588 | -0.379 | 0.016 | 2453796.4828 | 0.1747 | -0.399 | 0.016 |
| 2453796.6141 | 0.5885 | -1.095 | 0.045 | 2453794.5828 | 0.1888 | -0.815 | 0.033 | 2453796.5275 | 0.3158 | -0.445 | 0.018 |
| 2453796.6167 | 0.5969 | -1.163 | 0.048 | 2453794.5828 | 0.0638 | -0.403 | 0.017 | 2453796.5302 | 0.3241 | -0.437 | 0.018 |
| 2453796.6194 | 0.6053 | -1.207 | 0.049 | 2453794.5844 | 0.1938 | -0.835 | 0.034 | 2453796.5328 | 0.3324 | -0.427 | 0.017 |
| 2453796.6220 | 0.6136 | -1.180 | 0.048 | 2453794.5860 | 0.1989 | -0.843 | 0.034 | 2453796.5355 | 0.3408 | -0.417 | 0.017 |
| 2453796.6247 | 0.6220 | -1.208 | 0.049 | 2453794.5876 | 0.2040 | -0.873 | 0.036 | 2453796.5381 | 0.3492 | -0.414 | 0.017 |
| 2453796.6273 | 0.6303 | -1.227 | 0.050 | 2453794.5892 | 0.2090 | -0.863 | 0.035 | 2453796.5408 | 0.3575 | -0.392 | 0.016 |
| 2453796.6300 | 0.6387 | -1.234 | 0.050 | 2453794.5908 | 0.2141 | -0.883 | 0.036 | 2453796.5434 | 0.3658 | -0.387 | 0.016 |
| 2453796.6326 | 0.6470 | -1.256 | 0.051 | 2453794.5924 | 0.2192 | -0.887 | 0.036 | 2453796.5461 | 0.3742 | -0.372 | 0.015 |
| 2453796.6353 | 0.6554 | -1.287 | 0.053 | 2453794.5940 | 0.2242 | -0.845 | 0.035 | 2453796.5487 | 0.3826 | -0.347 | 0.014 |
| 2453796.6379 | 0.6638 | -1.276 | 0.052 | 2453794.6005 | 0.2445 | -0.904 | 0.037 | 2453796.5514 | 0.3909 | -0.330 | 0.014 |
| 2453796.6406 | 0.6721 | -1.279 | 0.052 | 2453794.6069 | 0.2648 | -0.894 | 0.037 | 2453796.5540 | 0.3992 | -0.306 | 0.013 |
| 2453796.6432 | 0.6804 | -1.307 | 0.053 | 2453794.6117 | 0.2801 | -0.890 | 0.036 | 2453796.5567 | 0.4076 | -0.277 | 0.011 |
| 2453796.6459 | 0.6888 | -1.326 | 0.054 | 2453794.6133 | 0.2851 | -0.875 | 0.036 | 2453796.5593 | 0.4160 | -0.255 | 0.010 |
| 2453796.6485 | 0.6972 | -1.320 | 0.054 | 2453796.4334 | 0.0168 | -0.218 | 0.009 | 2453796.5620 | 0.4243 | -0.222 | 0.009 |
| 2453796.6512 | 0.7055 | -1.317 | 0.054 | 2453796.4360 | 0.0252 | -0.237 | 0.01 | 2453796.5646 | 0.4327 | -0.181 | 0.007 |
| 2453796.6597 | 0.7326 | -1.327 | 0.054 | 2453796.4387 | 0.0335 | -0.303 | 0.012 | 2453796.5673 | 0.4410 | -0.154 | 0.006 |
| 2453911.3761 | 0.4437 | -0.999 | 0.041 | 2453796.4413 | 0.0419 | -0.346 | 0.014 | 2453796.5699 | 0.4494 | -0.115 | 0.005 |
| 2453911.3793 | 0.4538 | -0.949 | 0.039 | 2453796.4439 | 0.0502 | -0.413 | 0.017 | 2453796.5726 | 0.4577 | -0.076 | 0.003 |
| 2453911.3821 | 0.4627 | -0.924 | 0.038 | 2453796.4466 | 0.0585 | -0.455 | 0.019 | 2453796.5752 | 0.4660 | -0.029 | 0.001 |
| 2453911.3849 | 0.4716 | -0.892 | 0.036 | 2453796.4492 | 0.0669 | -0.514 | 0.021 | 2453796.5778 | 0.4744 | -0.001 | 0.000 |
| 2453911.3878 | 0.4805 | -0.841 | 0.034 | 2453796.4519 | 0.0752 | -0.550 | 0.023 | 2453796.5805 | 0.4828 | 0.031 | 0.001 |
| 2453911.3906 | 0.4894 | -0.796 | 0.033 | 2453796.4545 | 0.0836 | -0.589 | 0.024 | 2453796.5831 | 0.4911 | 0.045 | 0.002 |
| 2453911.3934 | 0.4984 | -0.778 | 0.032 | 2453796.4572 | 0.0920 | -0.608 | 0.025 | 2453796.5858 | 0.4994 | 0.054 | 0.002 |
| 2453911.3963 | 0.5073 | -0.829 | 0.034 | 2453796.4598 | 0.1003 | -0.641 | 0.026 | 2453796.5884 | 0.5077 | 0.044 | 0.002 |
| 2453911.3991 | 0.5162 | -0.828 | 0.034 | 2453796.4625 | 0.1087 | -0.673 | 0.028 | 2453796.5911 | 0.5161 | 0.028 | 0.001 |
| 2453911.4019 | 0.5252 | -0.863 | 0.035 | 2453796.4652 | 0.1170 | -0.705 | 0.029 | 2453796.5937 | 0.5245 | 0.006 | 0.000 |
| 2453911.4048 | 0.5341 | -0.863 | 0.035 | 2453796.4678 | 0.1254 | -0.715 | 0.029 | 2453796.5964 | 0.5328 | -0.021 | 0.001 |
| 2453911.4076 | 0.5430 | -0.941 | 0.038 | 2453796.4704 | 0.1337 | -0.737 | 0.03 | 2453796.5990 | 0.5412 | -0.058 | 0.002 |
| 2453911.4104 | 0.5519 | -0.964 | 0.039 | 2453796.4731 | 0.1421 | -0.749 | 0.031 | 2453796.6017 | 0.5496 | -0.090 | 0.004 |
| 2453911.4133 | 0.5609 | -1.003 | 0.041 | 2453796.4757 | 0.1504 | -0.762 | 0.031 | 2453796.6043 | 0.5579 | -0.131 | 0.005 |
| 2453911.4161 | 0.5698 | -1.063 | 0.043 | 2453796.4784 | 0.1588 | -0.783 | 0.032 | 2453796.6070 | 0.5663 | -0.169 | 0.007 |
| 2453911.4189 | 0.5788 | -1.093 | 0.045 | 2453796.4810 | 0.1671 | -0.799 | 0.033 | 2453796.6097 | 0.5747 | -0.204 | 0.008 |
| 2453911.4218 | 0.5878 | -1.122 | 0.046 | 2453796.4837 | 0.1754 | -0.800 | 0.033 | 2453796.6123 | 0.5830 | -0.239 | 0.010 |
| 2453911.4246 | 0.5967 | -1.153 | 0.047 | 2453796.4863 | 0.1838 | -0.819 | 0.033 | 2453796.6149 | 0.5914 | -0.271 | 0.011 |
| 2453911.4275 | 0.6057 | -1.174 | 0.048 | 2453796.4890 | 0.1922 | -0.854 | 0.035 | 2453796.6176 | 0.5997 | -0.301 | 0.012 |
| 2453911.4303 | 0.6145 | -1.177 | 0.048 | 2453796.4916 | 0.2005 | -0.855 | 0.035 | 2453796.6203 | 0.6081 | -0.318 | 0.013 |
| 2453911.4331 | 0.6235 | -1.236 | 0.051 | 2453796.4943 | 0.2089 | -0.850 | 0.035 | 2453796.6229 | 0.6165 | -0.340 | 0.014 |
| 2453911.4360 | 0.6325 | -1.249 | 0.051 | 2453796.4969 | 0.2173 | -0.854 | 0.035 | 2453796.6255 | 0.6248 | -0.358 | 0.015 |
| 2453911.4388 | 0.6414 | -1.282 | 0.052 | 2453796.4996 | 0.2257 | -0.868 | 0.035 | 2453796.6282 | 0.6331 | -0.375 | 0.015 |
| 2453911.4416 | 0.6504 | -1.264 | 0.052 | 2453796.5023 | 0.2340 | -0.880 | 0.036 | 2453796.6309 | 0.6415 | -0.391 | 0.016 |
| 2453911.4456 | 0.6629 | -1.332 | 0.054 | 2453796.5049 | 0.2424 | -0.890 | 0.036 | 2453796.6335 | 0.6499 | -0.404 | 0.017 |
| 2453911.4484 | 0.6718 | -1.318 | 0.054 | 2453796.5076 | 0.2507 | -0.885 | 0.036 | 2453796.6362 | 0.6582 | -0.418 | 0.017 |
| 2453911.4513 | 0.6807 | -1.336 | 0.055 | 2453796.5102 | 0.2591 | -0.885 | 0.036 | 2453796.6388 | 0.6666 | -0.421 | 0.017 |
| 2453911.4541 | 0.6896 | -1.347 | 0.055 | 2453796.5129 | 0.2675 | -0.888 | 0.036 | 2453796.6415 | 0.6749 | -0.440 | 0.018 |
| 2453911.4569 | 0.6985 | -1.319 | 0.054 | 2453796.5155 | 0.2758 | -0.882 | 0.036 | 2453796.6441 | 0.6832 | -0.442 | 0.018 |
| 2453911.4598 | 0.7075 | -1.319 | 0.054 | 2453796.5284 | 0.3165 | -0.858 | 0.035 | 2453796.6467 | 0.6916 | -0.458 | 0.019 |
| 2453911.4626 | 0.7164 | -1.315 | 0.054 | 2453796.5311 | 0.3249 | -0.834 | 0.034 | 2453796.6494 | 0.7000 | -0.460 | 0.019 |
| 2453911.4654 | 0.7254 | -1.332 | 0.054 | 2453796.5337 | 0.3332 | -0.832 | 0.034 | 2453796.6521 | 0.7084 | -0.471 | 0.019 |
| 2453911.4682 | 0.7343 | -1.315 | 0.054 | 2453796.5364 | 0.3416 | -0.823 | 0.034 | 2453835.3908 | 0.8453 | -0.363 | 0.015 |
| 2453911.4711 | 0.7432 | -1.329 | 0.054 | 2453796.5390 | 0.3500 | -0.818 | 0.0334 | 2453835.3938 | 0.8546 | -0.347 | 0.014 |
| 2453911.4711 | 0.7432 | -1.329 | 0.054 | 2453796.5417 | 0.3583 | -0.802 | 0.033 | 2453835.3975 | 0.8782 | -0.325 | 0.013 |
| 2453911.4739 | 0.7522 | -1.340 | 0.055 | 2453796.5443 | 0.3667 | -0.783 | 0.032 | 2453835.4012 | 0.8899 | -0.282 | 0.012 |
| 2453911.4739 | 0.7522 | -1.340 | 0.055 | 2453796.5470 | 0.3750 | -0.768 | 0.031 | 2453835.4049 | 0.9015 | -0.244 | 0.010 |
| 2453911.4768 | 0.7611 | -1.369 | 0.056 | 2453796.5496 | 0.3833 | -0.752 | 0.031 | 2453835.4086 | 0.9132 | -0.198 | 0.008 |
| 2453911.4796 | 0.7700 | -1.376 | 0.056 | 2453796.5523 | 0.3917 | -0.728 | 0.03 | 2453835.4123 | 0.9248 | -0.146 | 0.006 |
| 2453911.4851 | 0.7873 | -1.323 | 0.054 | 2453796.5549 | 0.4000 | -0.714 | 0.029 | 2453835.4159 | 0.9361 | -0.083 | 0.003 |
| 2453911.4879 | 0.7963 | -1.335 | 0.055 | 2453796.5575 | 0.4083 | -0.675 | 0.028 | 2453835.4196 | 0.9479 | -0.033 | 0.001 |
| 2453911.4908 | 0.8052 | -1.265 | 0.052 | 2453796.5602 | 0.4168 | -0.648 | 0.027 | 2453835.4233 | 0.9595 | 0.042 | 0.002 |
| 2453911.4936 | 0.8141 | -1.253 | 0.051 | 2453796.5629 | 0.4251 | -0.616 | 0.025 | 2453835.4270 | 0.9712 | 0.096 | 0.004 |
| 2453911.4964 | 0.8230 | -1.263 | 0.052 | 2453796.5655 | 0.4334 | -0.567 | 0.023 | 2453835.5332 | 0.3040 | -0.464 | 0.019 |
| 2453911.4992 | 0.8319 | -1.274 | 0.052 | 2453796.5682 | 0.4418 | -0.545 | 0.022 | 2453835.5369 | 0.3157 | -0.450 | 0.018 |
| 2453911.5021 | 0.8409 | -1.281 | 0.052 | 2453796.5708 | 0.4501 | -0.506 | 0.021 | 2453835.5406 | 0.3274 | -0.435 | 0.018 |
| 2453911.5049 | 0.8498 | -1.200 | 0.049 | 2453796.5734 | 0.4585 | -0.459 | 0.019 | 2453835.5443 | 0.3390 | -0.418 | 0.017 |
| 2453911.5077 | 0.8587 | -1.214 | 0.050 | 2453796.5761 | 0.4668 | -0.441 | 0.018 | 2453835.5480 | 0.3507 | -0.402 | 0.016 |
| 2453911.5105 | 0.8676 | -1.182 | 0.048 | 2453796.5787 | 0.4751 | -0.385 | 0.016 | 2453835.5516 | 0.3623 | -0.389 | 0.016 |
| 2453911.5133 | 0.8764 | -1.138 | 0.047 | 2453796.5814 | 0.4835 | -0.365 | 0.015 | 2453835.5554 | 0.3740 | -0.374 | 0.015 |
| 2453911.5162 | 0.8854 | -1.147 | 0.047 | 2453796.5840 | 0.4919 | -0.352 | 0.014 | 2453835.5591 | 0.3857 | -0.352 | 0.014 |
| 2453911.5190 | 0.8943 | -1.149 | 0.047 | 2453796.5893 | 0.5085 | -0.361 | 0.015 | 2453835.5627 | 0.3973 | -0.319 | 0.013 |
| 2453911.5219 | 0.9033 | -1.087 | 0.044 | 2453796.5920 | 0.5169 | -0.377 | 0.015 | 2453850.5833 | 0.7585 | -0.442 | 0.018 |
| 2453911.5247 | 0.9122 | -1.069 | 0.044 | 2453796.5946 | 0.5253 | -0.412 | 0.017 | 2453850.5860 | 0.7669 | -0.43 | 0.018 |
| 2453911.5275 | 0.9211 | -1.046 | 0.043 | 2453796.5973 | 0.5337 | -0.437 | 0.018 | 2453850.5886 | 0.7753 | -0.437 | 0.018 |
| 2453911.5303 | 0.9301 | -0.966 | 0.039 | 2453796.5999 | 0.5420 | -0.469 | 0.019 | 2453850.5939 | 0.7920 | -0.433 | 0.018 |
| 2453911.5332 | 0.9390 | -0.963 | 0.039 | 2453796.6026 | 0.5504 | -0.499 | 0.02 | 2453850.5966 | 0.8004 | -0.427 | 0.017 |
| 2453911.5360 | 0.9479 | -0.863 | 0.035 | 2453796.6052 | 0.5587 | -0.551 | 0.023 | 2453850.5993 | 0.8088 | -0.426 | 0.017 |
| 2453911.5388 | 0.9568 | -0.780 | 0.032 | 2453796.6079 | 0.5671 | -0.585 | 0.024 | 2453850.6019 | 0.8172 | -0.410 | 0.017 |
| 2453911.5417 | 0.9658 | -0.698 | 0.029 | 2453796.6105 | 0.5755 | -0.629 | 0.026 | 2453850.6046 | 0.8256 | -0.404 | 0.017 |
| 2453911.5445 | 0.9747 | -0.664 | 0.027 | 2453796.6132 | 0.5838 | -0.675 | 0.028 | 2453850.6072 | 0.8340 | -0.382 | 0.016 |
| 2453911.5474 | 0.9837 | -0.668 | 0.027 | 2453796.6158 | 0.5922 | -0.679 | 0.028 | 2453878.5321 | 0.8956 | -0.313 | 0.013 |
| 2453911.5502 | 0.9926 | -0.665 | 0.027 | 2453796.6185 | 0.6005 | -0.726 | 0.03 | 2453878.5344 | 0.9030 | -0.222 | 0.009 |
| 2453911.5530 | 0.0015 | -0.618 | 0.025 | 2453796.6211 | 0.6088 | -0.753 | 0.031 | 2453878.5371 | 0.9113 | -0.206 | 0.008 |
| 2453911.5559 | 0.0105 | -0.606 | 0.025 | 2453796.6238 | 0.6172 | -0.769 | 0.031 | 2453878.5397 | 0.9197 | -0.230 | 0.009 |
| 2453911.5587 | 0.0194 | -0.652 | 0.027 | 2453796.6264 | 0.6256 | -0.778 | 0.032 | 2453878.5530 | 0.9614 | 0.006 | 0.000 |
| 2453911.5615 | 0.0283 | -0.668 | 0.027 | 2453796.6291 | 0.6339 | -0.794 | 0.032 | 2453878.5556 | 0.9698 | 0.060 | 0.003 |
| 2453911.5643 | 0.0372 | -0.746 | 0.031 | 2453796.6317 | 0.6422 | -0.802 | 0.033 | 2453878.5583 | 0.9782 | 0.086 | 0.004 |
| 2453911.5672 | 0.0461 | -0.752 | 0.031 | 2453796.6344 | 0.6507 | -0.830 | 0.034 | 2453878.5609 | 0.9745 | 0.124 | 0.005 |
| 2453911.5700 | 0.0551 | -0.880 | 0.036 | 2453796.6370 | 0.6590 | -0.836 | 0.034 | 2453878.5636 | 0.9829 | 0.147 | 0.006 |
| 2453911.5728 | 0.0640 | -0.983 | 0.040 | 2453796.6397 | 0.6673 | -0.857 | 0.035 | 2453878.5662 | 0.9912 | 0.155 | 0.006 |
| 2453911.5756 | 0.0729 | -0.970 | 0.040 | 2453796.6423 | 0.6757 | -0.854 | 0.035 | 2453911.3733 | 0.8968 | -0.183 | 0.008 |
| 2453911.5785 | 0.0818 | -1.112 | 0.045 | 2453796.6450 | 0.6841 | -0.866 | 0.035 | 2453911.3774 | 0.9099 | -0.155 | 0.006 |
| 2453911.5813 | 0.0908 | -1.070 | 0.044 | 2453796.6476 | 0.6924 | -0.879 | 0.036 | 2453911.3803 | 0.9188 | -0.106 | 0.004 |
| 2453911.5841 | 0.0997 | -1.042 | 0.043 | 2453796.6503 | 0.7007 | -0.883 | 0.036 | 2453911.3831 | 0.9278 | -0.067 | 0.003 |
| 2453911.5870 | 0.1086 | -1.121 | 0.046 | 2453796.6589 | 0.7728 | -0.865 | 0.035 | 2453911.3859 | 0.9367 | -0.034 | 0.001 |
| 2453911.5898 | 0.1175 | -1.083 | 0.044 | 2453796.6615 | 0.7811 | -0.852 | 0.035 | 2453911.4437 | 0.6570 | -0.431 | 0.018 |
| 2453911.5926 | 0.1265 | -1.184 | 0.048 | 2453911.5264 | 0.9378 | -0.525 | 0.021 | 2453911.4466 | 0.6660 | -0.425 | 0.017 |
| 2453911.5955 | 0.1354 | -1.234 | 0.050 | 2453911.5293 | 0.9467 | -0.469 | 0.019 | 2453911.4522 | 0.6838 | -0.450 | 0.018 |
| 2453911.5983 | 0.1444 | -1.270 | 0.052 | 2453911.5321 | 0.9556 | -0.437 | 0.018 | 2453911.4579 | 0.7016 | -0.478 | 0.020 |
| 2453911.6012 | 0.1533 | -1.240 | 0.051 | 2453911.5349 | 0.9646 | -0.367 | 0.015 | 2453911.4607 | 0.7106 | -0.475 | 0.019 |
| 2453911.6040 | 0.1622 | -1.243 | 0.051 | 2453911.5378 | 0.9735 | -0.292 | 0.012 | 2453911.4636 | 0.7195 | -0.476 | 0.019 |
| 2453911.6068 | 0.1711 | -1.285 | 0.053 | 2453911.5406 | 0.9825 | -0.253 | 0.01 | 2453911.5455 | 0.9778 | 0.129 | 0.005 |

Table 2: Light minima of TY Boo in BVR band.

| HJD | Error | Min | Filter |
|---|---|---|---|
| 2453794.47770 | 0.0003 | I | B |
| 2453794.48280 | 0.0006 | I | V |
| 2453794.49470 | 0.0002 | I | R |
| 2453796.55590 | 0.0005 | II | B |
| 2453796.55630 | 0.0002 | II | V |
| 2453796.55640 | 0.0001 | II | R |
| 2453835.39900 | 0.0013 | I | B |
| 2453835.39902 | 0.0001 | I | V |
| 2453835.39903 | 0.0002 | I | R |
| 2453850.46370 | 0.0005 | I | V |
| 2453850.46670 | 0.0015 | II | B |
| 2453850.47210 | 0.0003 | II | V |
| 2453850.62220 | 0.0004 | II | R |
| 2453878.53750 | 0.0005 | I | V |
| 2453878.53780 | 0.0011 | I | R |
| 2453911.36400 | 0.0004 | I | B |
| 2453911.52300 | 0.0006 | I | V |
| 2453911.52360 | 0.0004 | I | R |
| 2453911.52370 | 0.0010 | II | B |

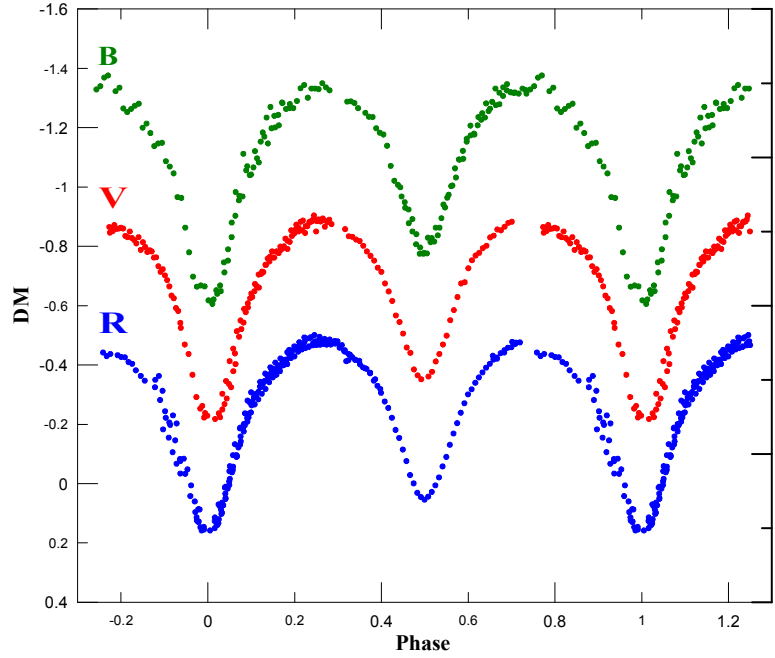

Figure 1: BVR light curves of TY Boo.

## 3. Orbital Period Analysis

The orbital period stability of the system TY Boo was first recognized by Szafraniec (1953), who found a cyclic period variation of about 127 days. A cubic ephemeris was obtained by Wood&Forbes (1963) with a rate of period increase of dp/dE =+7.79 x $10^{-10}$ days/cycle (dp/dt = +1.79 x $10^{-6}$ days/yr). Their study based on only 22 light of minima. Carr (1972) confirms the period (p=0.317146) given by Wood&Forbes (1963).

A period study by Samec and Bookmer (1987) gives no indication of the cyclic period variation which suggested by Szafraniec. A non continuous period variation was announced by Milone et al. (1991). They recorded a rapid developing Ca II flares in the system. Li et al. (2005) found two periodical variations (31.5 and 11.76 yr) superimpose on a continuous increase (dp/dt = +6.28 x $10^{-8}$ days/yr). They refer the continuous increase to the mass transfer from the secondary to the primary, rather than an expansion of the primary due to its dynamical instability. Yang et al. (2007) concluded a secular period decrease superimposed on a cyclic variation. They attributed this decrease to the mass transfer from the more massive component to the less massive component, which appear in shrinking of the inner and outer critical Roche lobes, and then caused the contact degree to increase. A period study based on published minima until 2011 by Christopoulou et al (2012) showed a long term period decrease.

In the present work we used the list collected by Christopoulou et al. which cover the interval from 1926 to 2011 together with our new 19 minima and all published from 2011 to 2014. In addition we added published minima before 2011 and have not included in Christopoulou et al. list. A total of 76 minima were added in our study, these results in a total of 454 light minimum timings, covering over 88 years (~ 101277 revolutions) were used to follow and update the long term orbital behavior of the system TY Boo by means of an (O-C) diagram. The C's values of the eclipse timings have been calculated using the linear ephemeris (Kreiner et al. 2001) (Eq. 1) and listed in Table 3.

For more accurate result, we discarded some collected uncertain minimum times (i.e. the first visual minimum of 1926) which are not in harmony with others in the (O-C) diagram. Scattering of the minima in the (O-C) diagram may be results from cycle to cycle variation in the observed light curves, which leads to non-symmetry and also uncertainty in times of minima calculations. The (O-C) values were represented in Figure 2 versus the integer cycle E, no distinctions have been made between primary and secondary minima. It's clear that, the behavior of the (O-C) points in Figure 2 can't represent any mean elements derived by linear best fit. In order to follow the period behavior for the system TY Boo though the 88 years since its discovery, we divided the (O-C) variation into four intervals, **$E_i - E_{i-1}$,** i=1,2,3,4.

Table 4 summarized the intervals and the best fit data with standard deviations SD, correlation coefficients **r** and the residual sum of squares. The time interval ΔE and the corresponding changes in the period ΔP, for each interval according to the best fit of the (O-C) residuals are also listed in Table 4.

It's clear from the table that the period of the system TY Boo shows two stages of increase and similar of decrease, which looks like as a periodic behavior. The (O-C) residuals in Figure 2 shows two peaks, represented the turning points from phase of period increase to decrease. The first peak at HJD ~ 24329961949 (1949) while the second one at HJD ~ 2452362 (2002), with interval of about 53 yr between them. The general trend of the (O-C) diagram can represent by a six degree polynomial with a residual sum of square = 0.0094 and correlation coefficients = 0.89, as following:

$$\text{Min } I = 2447612.6040 + 0.3171506 \cdot E - 2.929 \times 10^{-12} \cdot E^2 - 2.051 \times 10^{-15} \cdot E^3$$
$$- 7.491 \times 10^{-21} \cdot E^4 + 4.049 \times 10^{-25} \cdot E^5 - 9.218 \times 10^{-31} \cdot E^6 \qquad (2)$$

The new light elements of Eq. 2 can be used to estimate minimum times in the next few years. The equation yields a new period (p= $0.^d3171506$). The period shows a decrease with the rate dP/dE = 5.858 $\times 10^{-12}$ day/cycle or 6.742 $\times 10^{-9}$ day/year or 0.058 second/century. The (O–C)$p$ residuals calculated using polynomial ephemeris (Eq. 2) are listed in Table 3 and displayed in Figure 3.

Table 3: Times of minimum light for TY Boo (2011-2014.)

| HJD | Method | E | O-C | (O-C)p | References |
|---|---|---|---|---|---|
| 2432688.5390 | vis | -47057 | 0.0160 | 0.0038 | Szafraniec (1948) |
| 2432688.5390 | vis | -47057 | 0.0160 | 0.0038 | Szafraniec (1948) |
| 2433082.4350 | vis | -45815 | 0.0129 | 0.0009 | Szafraniec (1948) |
| 2433362.4660 | vis | -44932 | 0.0014 | -0.0102 | Szafraniec (1948) |
| 2443587.8020 | vis | -12690.5 | -0.0221 | -0.0057 | Samolyk (1992) |
| 2444334.7090 | vis | -10335.5 | -0.0010 | 0.0132 | Samolyk (1992) |
| 2444402.7250 | vis | -10121 | -0.0135 | 0.0005 | Samolyk (1992) |
| 2445173.7150 | vis | -7690 | -0.0127 | -0.0016 | Samolyk (1992) |
| 2445492.7770 | vis | -6684 | -0.0026 | 0.0072 | Samolyk (1992) |
| 2446210.7980 | vis | -4420 | -0.0069 | -0.0005 | Samolyk (1992) |
| 2446210.8000 | vis | -4420 | -0.0049 | 0.0015 | Samolyk (1992) |
| 2446226.8110 | Pe | -4369.5 | -0.0100 | -0.0036 | Milon et al (1991) |
| 2446227.7629 | Pe | -4366.5 | -0.0095 | -0.0031 | Milon et al (1991) |
| 2446231.7266 | Pe | -4354 | -0.0102 | -0.0038 | Milon et al (1991) |
| 2446600.7230 | vis | -3190.5 | -0.0166 | -0.0121 | Samolyk (1992) |
| 2446606.7660 | vis | -3171.5 | 0.0006 | 0.0051 | Samolyk (1992) |
| 2446678.6020 | vis | -2945 | 0.0023 | 0.0065 | Samolyk (1992) |
| 2446875.7160 | vis | -2323.5 | 0.0082 | 0.0114 | Samolyk (1992) |
| 2446951.6690 | vis | -2084 | 0.0040 | 0.0069 | Samolyk (1992) |
| 2447010.6550 | vis | -1898 | 0.0003 | 0.0028 | Samolyk (1992) |
| 2447219.8060 | vis | -1238.5 | -0.0085 | -0.0070 | Samolyk (1992) |
| 2447263.5774 | vis | -1100.5 | -0.0036 | -0.0024 | Agerer (1988) |
| 2447263.5781 | vis | -1100.5 | -0.0029 | -0.0017 | Agerer (1988) |
| 2447299.7430 | vis | -986.5 | 0.0070 | 0.0081 | Samolyk (1992) |
| 2447316.7060 | vis | -933 | 0.0025 | 0.0035 | Samolyk (1992) |
| 2447612.6025 | vis | 0 | -0.0010 | -0.0015 | Hubscher et al. (1989) |

| JD | Type | E | O-C1 | O-C2 | Reference |
|---|---|---|---|---|---|
| 2447612.6032 | vis | 0 | -0.0003 | -0.0008 | Hubscher et al. (1989) |
| 2447681.7510 | vis | 218 | 0.0090 | 0.0082 | Samolyk (1992) |
| 2448161.5940 | vis | 1731 | 0.0056 | 0.0023 | Samolyk (1992) |
| 2448330.6410 | vis | 2264 | 0.0122 | 0.0081 | Samolyk (1992) |
| 2448331.7410 | vis | 2267.5 | 0.0021 | -0.0020 | Samolyk (1992) |
| 2448661.8970 | vis | 3308.5 | 0.0060 | 0.0004 | Samolyk (1992) |
| 2448690.5994 | vis | 3399 | 0.0065 | 0.0006 | Hubscher et al. (1992) |
| 2448717.7150 | vis | 3484.5 | 0.0058 | -0.0001 | Samolyk (1992) |
| 2448724.8440 | vis | 3507 | -0.0010 | -0.0070 | Samolyk (1992) |
| 2448770.6700 | vis | 3651.5 | -0.0031 | -0.0093 | Samolyk (1992) |
| 2453794.4777 | ccd | 19492 | 0.0059 | -0.0096 | This paper |
| 2453794.4828 | ccd | 19492 | 0.0110 | -0.0045 | This paper |
| 2453794.4947 | ccd | 19492 | 0.0229 | 0.0074 | This paper |
| 2453796.5559 | ccd | 19498.5 | 0.0226 | 0.0071 | This paper |
| 2453796.5563 | ccd | 19498.5 | 0.0230 | 0.0075 | This paper |
| 2453796.5564 | ccd | 19498.5 | 0.0231 | 0.0076 | This paper |
| 2453835.3990 | ccd | 19621 | 0.0150 | -0.0004 | This paper |
| 2453835.3990 | ccd | 19621 | 0.0150 | -0.0004 | This paper |
| 2453835.3990 | ccd | 19621 | 0.0150 | -0.0004 | This paper |
| 2453850.4637 | ccd | 19668.5 | 0.0151 | -0.0003 | This paper |
| 2453850.4667 | ccd | 19668.5 | 0.0181 | 0.0027 | This paper |
| 2453850.4721 | ccd | 19668.5 | 0.0235 | 0.0081 | This paper |
| 2453850.6222 | ccd | 19669 | 0.0150 | -0.0003 | This paper |
| 2453878.5375 | ccd | 19757 | 0.0212 | 0.0059 | This paper |
| 2453878.5378 | ccd | 19757 | 0.0215 | 0.0062 | This paper |
| 2453911.3640 | ccd | 19860.5 | 0.0228 | 0.0076 | This paper |
| 2453911.5230 | ccd | 19861 | 0.0232 | 0.0080 | This paper |
| 2453911.5236 | ccd | 19861 | 0.0238 | 0.0086 | This paper |
| 2453911.5237 | ccd | 19861 | 0.0239 | 0.0087 | This paper |
| 2454958.7319 | vis | 23163 | 0.0061 | -0.0051 | Bialozynski (2009) |
| 2454958.7420 | vis | 23163 | 0.0162 | 0.0050 | Bialozynski (2009) |
| 2455232.9082 | vis | 24027.5 | 0.0071 | -0.0026 | Menzies (2010) |
| 2455642.3477 | vis | 25318.5 | 0.0072 | 0.0000 | Agerer (2011) |
| 2455642.5067 | vis | 25319 | 0.0077 | 0.0004 | Agerer (2011) |
| 2455648.3732 | ccd | 25337.5 | 0.0069 | -0.0003 | Hübscher et al. (2012) |
| 2455664.0717 | ccd | 25387 | 0.0065 | -0.0006 | Shiokawa (2011) |
| 2455664.2305 | ccd | 25387.5 | 0.0068 | -0.0003 | Shiokawa (2011) |
| 2455681.5158 | vis | 25442 | 0.0074 | 0.0005 | Parimucha et al. (2013) |
| 2455992.6392 | ccd | 26423 | 0.0077 | 0.0029 | Hoonkova (2012) |
| 2455992.6395 | ccd | 26423 | 0.0080 | 0.0032 | Hoonkova (2012) |
| 2456023.8779 | vis | 26521.5 | 0.0072 | 0.0027 | Diethelm (2012) |
| 2456062.7300 | vis | 26644 | 0.0085 | 0.0043 | Sabo (2012) |
| 2456069.3859 | vis | 26665 | 0.0043 | 0.0001 | Hübscher et al. (2013) |
| 2456069.5470 | vis | 26665.5 | 0.0068 | 0.0026 | Hübscher et al. (2013) |
| 2456078.4252 | vis | 26693.5 | 0.0049 | 0.0007 | Parimucha et al. (2013) |
| 2456085.7218 | vis | 26716.5 | 0.0070 | 0.0030 | Diethelm (2012) |
| 2456382.7220 | vis | 27653 | -0.0028 | -0.0045 | Menzies (2013) |
| 2456399.6891 | ccd | 27706.5 | -0.0032 | -0.0048 | Poklar (2013) |
| 2456408.4154 | vis | 27734 | 0.0015 | 0.0002 | Hübscher (2013) |
| 2456408.5742 | vis | 27734.5 | 0.0018 | 0.0003 | Hübscher (2013) |

Table 4: Comprehensive period behavior for the system TY Boo

| Parameters | Intervals (2400000+) | | | |
|---|---|---|---|---|
| | $E_0$ to $E_1$ 29348 - 32996 | $E_1$ to $E_2$ 32996 – 40370 | $E_2$ to $E_3$ 40370 -52452 | $E_3$ to $E_4$ 52452 - 56408 |
| $\Delta E$ (day) | 3648 | 7374 | 12082 | 3956 |
| Period (day) | 0.3171519 | 0.3171476 | 0.3171502 | 0.3171477 |
| $\Delta P$ (day) | $2.858 \times 10^{-6}$ | $-1.3949 \times 10^{-6}$ | $1.1849 \times 10^{-6}$ | $-1.2620 \times 10^{-6}$ |
| $\Delta P/P$ | $9.012 \times 10^{-6}$ | $-4.398 \times 10^{-6}$ | $3.7360 \times 10^{-6}$ | $-3.9790 \times 10^{-6}$ |
| $\Delta P/\Delta E$ (d/cycle) | $2.485 \times 10^{-10}$ | $-5.999 \times 10^{-11}$ | $3.1103 \times 10^{-11}$ | $-1.0120 \times 10^{-10}$ |
| Epoch (2400000+) | 47612.7507 | 47612.5531 | 47612.6040 | 47612.5643 |
| SD | 0.00559 | 0.00624 | 0.00503 | 0.00303 |
| r | 0.94300 | 0.85700 | 0.84700 | 0.82620 |
| Residual sum of square | 0.00041 | 0.00261 | 0.01134 | 0.00172 |
| R_Squared | 0.88980 | 0.73450 | 0.71720 | 0.68260 |

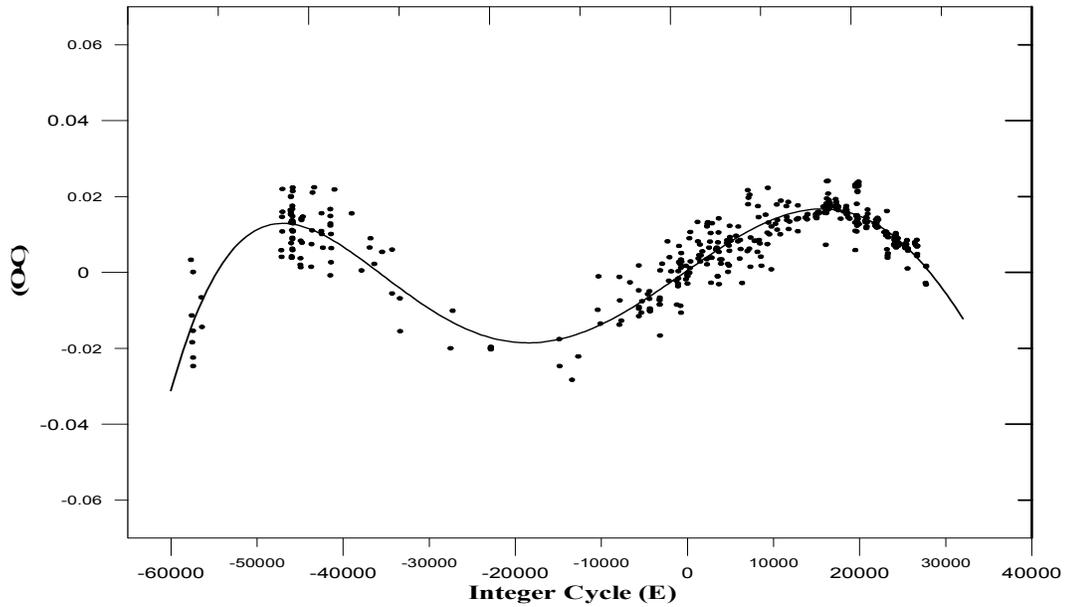

Figure 2: Period behavior of TY Boo.

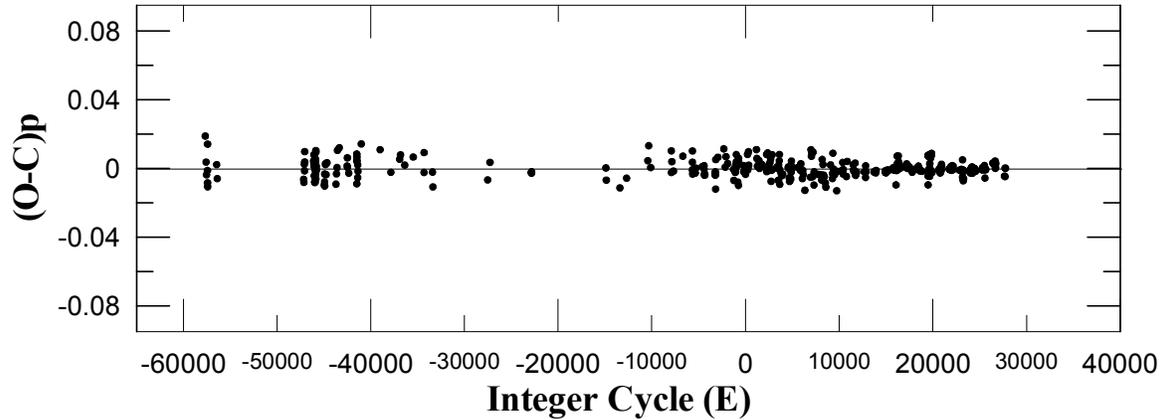
Figure 3: Calculated residuals from the quadratic ephemeris.

## 4. Light Curve Stability

The light curve variation of the system TY Boo was noted through the historical survey of the published light curves since its discovery in 1926. Observations by Carr (1972) in 1969 were brighter than that of Samec and Bookmyer (1987). Li et al. (2005) suggested that the more massive component is brighter in 1969 than in 1986, which may cause by star spot(s) or as a result of the stellar activity. A large scattering were clearly seen in the light curves reported by Samec and Bookmyer (1987) and Milone et al. (1991), and a rapid light variation night by night were observed near the two maxima and two minima. This phenomenon has been observed in many contact binaries (i.e. KN Per (Goderaya et al. 1997), CN And (Keskin 1989), and AQ Tuc Hilditch, King (1986)), which may cause by the pulsation of a common envelope due to mass transfer between two components (Li et al. 2002).

Studying of the light curve variation together with orbital period of W UMa systems is very important in understanding the evolution structure of these systems. Appliegate (1992) predicts a relation between the orbital period changes and light variations during the same cycle. We studied the possibility of applied Applegate prediction to the system TY Boo. Using the historical published light curves together with our observations in V band, the light levels (Max I, Min I, Max II, and Min II) were evaluated from the curves. The magnitude differences between both maxima (O'Connell) $D_{max}$ (Max I – Max II) and minima $D_{min}$ (Min I – Min II) and amplitude (depths) of the primary $A_p$ (Min I – Max I) and secondary $A_s$ (Min II – Max I) minima, have been calculated for each light curve and listed in Table 5 with their corresponding observers and observational date in years. Figure 4 displays the variation of the magnitude differences $D_{max}$, $D_{min}$, and the amplitude (depths) of the primary eclipse $A_p$ and secondary one $A_s$, for TY Boo in V band. From Figure 4a, c, and d we can note that the amplitude of the primary and secondary eclipse $A_p$, and $A_s$, showed together the same trend of variation, while an opposite trend was shown by $D_{max}$.

The tabulated results leads to a conclusion that the calculated parameters, $D_{max}$, $D_{min}$, $A_p$, and $A_s$, showed an oscillatory variation and wave like behavior as a period function, which was interpreted by a periodic action of some physical mechanism. Synchronous of periodic variation for both orbital period and light curves parameters ($D_{max}$, $D_{min}$, $A_p$, and $A_s$) for the system TY Boo may be interpret by the presence of magnetic activity cycles, and/or mass transfer mechanism.

Table 5: Light curve parameters for TY Boo

| HJD | Date | $D_{max}$ (mag) | $D_{min}$ (mag) | $A_p$ (mag) | $A_s$ (mag) | Ref. |
|---|---|---|---|---|---|---|
| 2441351 | 1972 | 0.009±0.001 | -0.119±0.006 | 0.357±0.018 | 0.476±0.024 | Carr (1972) |
| 2446830 | 1987 | -0.006±0.001 | 0.146±0.007 | 0.674±0.028 | 0.398±0.020 | Samec&Bookmye (1987) |
| 2447561 | 1989 | 0.005±0.001 | -0.070±0.004 | 0.455±0.023 | 0.385±0.019 | Samec et al (1989) |
| 2447926 | 1990 | 0.200±0.010 | 0.160±0.008 | 0.180±0.009 | 0.340±0.017 | Rainger et al. (1990) |
| 2448291 | 1991 | 0.030±0.002 | 0.075±0.004 | 0.460±0.023 | 0.385±0.019 | Millon et al. (1991) |
| 2453770 | 2006 | 0.010±0.001 | 0.135±0.006 | 0545±0.065 | 0.680±0.028 | This paper |
| 2454866 | 2009 | -- | 0.089±0.005 | -- | -- | Bialozynski (2009) |
| 2455596 | 2011 | 0.050±0.002 | 0.211±0.010 | 0.768±0.031 | 0557±0.023 | Christopolou et al. (2011) |
| 2456326 | 2013 | -- | 0.049±0.003 | -- | -- | Menzies (2013) |

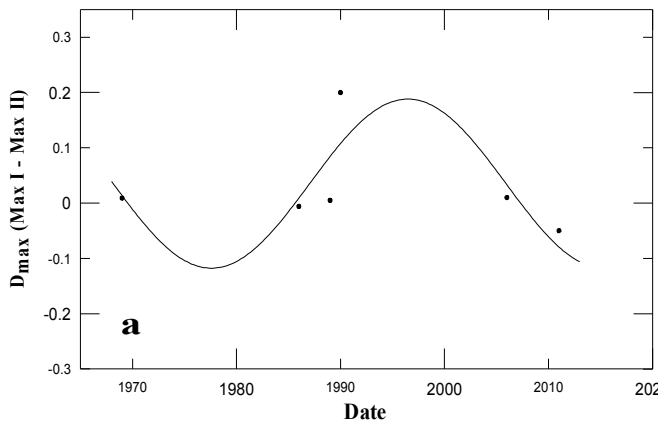

a

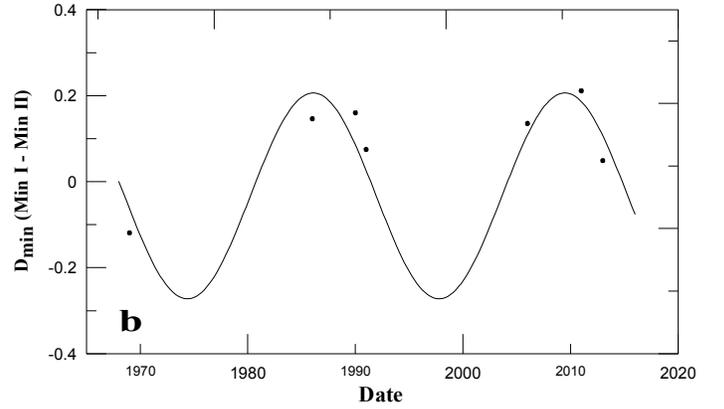

b

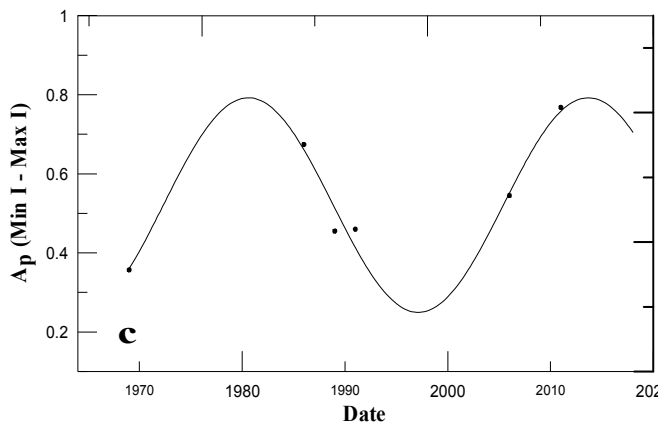

c

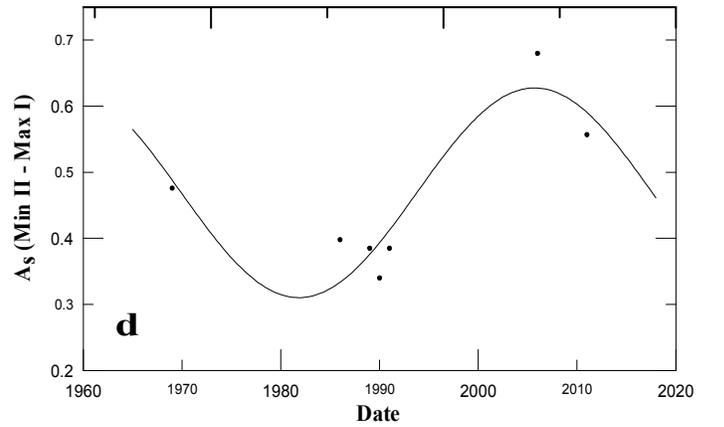

d

Figure 4: Variation of the magnitude differences $D_{max}$, $D_{min}$, and the amplitude (depths) of the primary $A_p$ and secondary $A_s$ for TY Boo in V band.

## 5. Light Curve Modeling

Light curve solutions for the W UMa system TY Boo estimated by many authors (i.e. Rainger et al. 1990, Milone et al 1991, and Christopoulou et al. 2012) showed that the less massive, but hotter, star is eclipsed at the primary minimum. Although the light curve solution by Carr (1972) has a limited accuracy, it suggested that the system TY Boo belong to A-type systems and consists of two main-sequence components with spectral types G3 and G7 and W UMa characteristic with mass ratio of 0.88. Niarchos (1978) re-analyzed Carr's (1972) observations using frequency domain techniques, and suggested that TY Boo was a W UMa system with a small mass ratio of 0.22 and indicated that the smaller component is the hotter one (Rainger et al. 1990).

Light curve analysis by Samec et al. (1989) shows that the system has a spectral type range between G4 and G8. Rainger et al. (1990) combined their spectroscopic observations together with photometric observations by Samec and Bookmyer (1987) to compute the masses and absolute dimensions for the components. Their solution shows that the system TY Boo has components of G2 and F8 spectral type. Milone et al. (1991) suggested a first spotted solution with both hot and cold spots on a cooler component, and showed that the published light curves by Carr (1972) and Samec and Bookmyer (1987) have a minimal asymmetry and can be fitted without spots.

Christopoulou et al. (2012) confirmed the spotted model solution suggested by Milone et al. (1990) and estimated a model included two-spot model on either stellar components. They suggested that the spots were caused by cyclic magnetic activity rather than an unseen companion. They compare the light curve parameters obtained from 1926 to 2011, which shows that the system TY Boo has been a difficult system to analyze because it yields widely divergent results in mass ratios and size.

Our observations cover three bandpass (BVR) and have high accuracy with respect to the previous published photometric data series. In our modeling we use Mode 3 (overcontact) of WDint56a Package (Nelson, 2009) which based on the Willson and Devinney (W-D) code. The surface temperature of the primary star (less massive and hotter star) was fixed at 5732 K, which is compatible with its spectral classification G3 (Cox, 2000).

The individual light curves observations were analyzed instead of normal light curves, which don't reveal a real light variation of the system. The bolometric limb darkening coefficients ($x_{b(h)} = x_{b(c)}$, $y_{b(h)} = y_{b(c)}$) were adopted and interpolated using the square-root law from Van Hamme (1993) Tables and model atmosphere were applied. Gravity darkening and bolometric albedo were assumed according to the exponents appropriate for convective envelopes ($T_{eff} < 7500$ K) of the late spectral type. We adopted $g_h = g_c = 0.32$ (Lucy, 1967) and the albedo value $A_h = A_c = 0.5$ (Rucinski, 1969). Mode 3 (overcontact) was applied with a synchronous rotation and circular orbit were assumed. Some parameters were kept fixed (i.e. $T_h$, g, A, X), and the adjusted parameters are the temperature of the cool star $T_c$ of the star 2, the monochromatic luminosity $L_1$ of star 1 (the luminosity of star 2 was calculated by the stellar atmosphere model), the surface potential $\Omega_h = \Omega_c$, and the mass ratio $q = (M_c/M_h)$. Model solution without spot (not shown here) doesn't fit the observed light curves well. The parameters of the accepted model are listed in Table 6 with a presence of dark spot on the cooler component and a hot one on the hotter, which confirms the spotted solution suggested by Christopoulou et al. (2012) and gave a credible uniform description for the system TY boo system. The estimated parameters show that the low massive component is hotter than the more massive and cool one, which confirmed the results of Milone et al. (1991) and Christopoulou et al. (2012). The temperature difference between the components is $\Delta T \sim 249\ ^0K$, while the theoretical BVR light curves for the system TY Boo are displaying in Figure 5. The absolute physical parameters of the system TY Boo were calculated based on the results of the radial velocity data of Milone et al. (1991) and our new photometric solution. The calculated parameters together with that calculated by previous light curve solutions were listed in Table 7. A three dimensional geometrical structure for the system TY Boo was displayed in Figure 6 using the software Package Binary Maker 3.03 (Bradstreet and Steelman, 2004) based on the calculated parameters resulted from our model. It's clear from our results that the system TY Boo shows a rapid transformation in the physical properties which have many interpretations, i.e. Milone et al. (1990) suggested that the system is Chromospherically active and a rapidly developing Ca II flare was recorded in the system's spectrum. A shrinking in the inner and outer critical Roche lobes caused an increasing in the contact degree as a result of the decrease of orbital period.

Table 6: Photometric solution for TY Boo.

| Parameter | BVR |
|---|---|
| $i(^o)$ | 78.76±0.16 |
| $g_h = g_c$ | Fixed |
| $A_h = A_c$ | Fixed |
| $q = (M_c / M_h)$ | 2.2592±0.005 |
| $\Omega_h = \Omega_c$ | 5.5935±0.010 |
| $\Omega_{in}$ | 5.6174 |
| $\Omega_{out}$ | 4.5571 |
| $T_h$ (K) | 5732 Fixed |
| $T_c$ (K) | 5483±2 |
| $L_h/(L_h+L_c)$ | Fixed |
| $L_c/(L_h+L_c)$ | Fixed |
| $r_h$ (pole) | 0.2920±0.0006 |
| $r_h$ (side) | 0.3048±0.0007 |
| $r_h$ (back) | 0.3385±0.0013 |
| $r_c$ (pole) | 0.4260±0.0002 |
| $r_c$ (side) | 0.4540±0.0003 |
| $r_c$ (back) | 0.4819±0.0006 |

*Spot parameters for hot star*

| | | |
|---|---|---|
| Co-latitude | 120 | Fixed |
| Longitude | 350 | Fixed |
| Spot radius | 16 | Fixed |
| Temp. factor | 1.1076 | Fixed |

*Spot parameters for hot star*

| | | |
|---|---|---|
| Co-latitude | 140 | Fixed |
| Longitude | 155 | Fixed |
| Spot radius | 21 | Fixed |
| Temp. factor | 0.7 | Fixed |
| $\sum (O-C)^2$ | 0.08447 | |

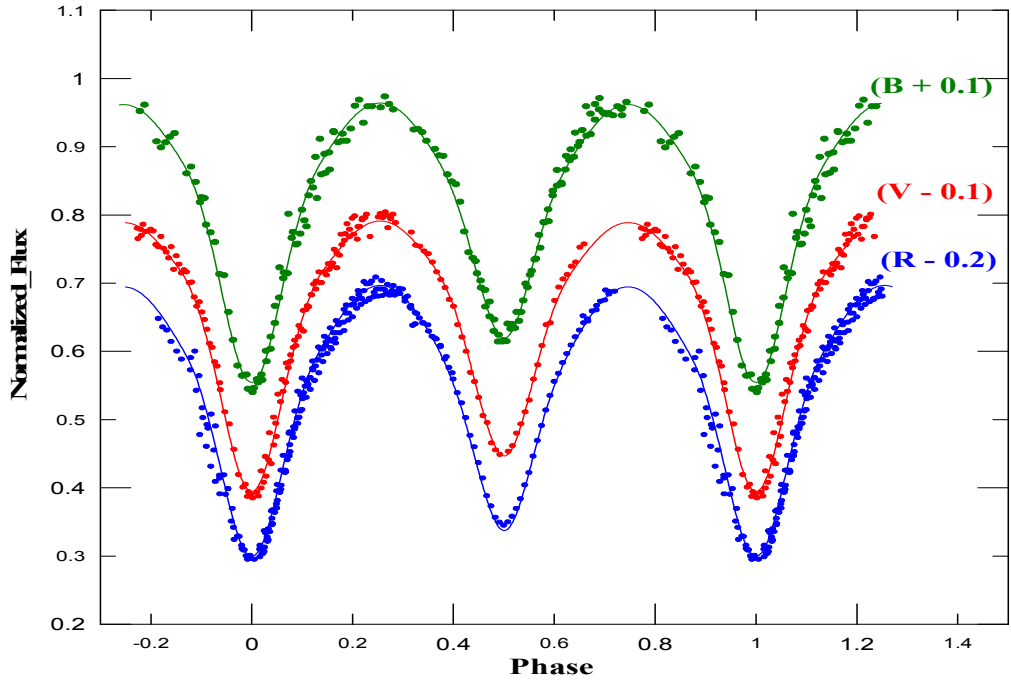

Figure 5: Observed light curves (filled circles), and the synthetic curves.

Table 7: Absolute physical parameters for the system TY Boo.

| Parameter | | | | | | | | | | Ref. |
|---|---|---|---|---|---|---|---|---|---|---|
| $M_h(M_\odot)$ | $M_c(M_\odot)$ | $R_h(R_\odot)$ | $R_c(R_\odot)$ | $M_{h\_bol}$ | $M_{c\_bol}$ | $L_h(L_\odot)$ | $L_c(L_\odot)$ | $T_h(T_\odot)$ | $T_c(T_\odot)$ | |
| 0.40±0.01 | 0.93±0.02 | 0.69±0.01 | 1.00±0.01 | 5.28±0.14 | 4.75±0.15 | 0.62±0.03 | 1.02±0.04 | 1.07±0.04 | 1.00±0.04 | 1 |
| 0.53±0.02 | 1.14±0.05 | 0.75±0.03 | 1.05±0.04 | 5.29±0.22 | 4.82±0.02 | 0.58±0.02 | 0.89±0.04 | 1.01±0.04 | 0.94±0.04 | 2 |
| 0.57±0.05 | 1.21±0.06 | 0.75±0.01 | 1.07±0.01 | 5.34±0.22 | 4.88±0.02 | 0.54±0.01 | 0.87±0.02 | 0.99±0.04 | 0.88±0.04 | 3 |
| 0.53±0.02 | 1.19±0.05 | 0.73±0.03 | 1.06±0.04 | 5.47±0.22 | 4.85±0.20 | 0.52±0.02 | 0.92±0.04 | 0.99±0.04 | 0.95±0.04 | 4 |

\* Subscript **h** and **c** means hot and cool component respectively.
**Reference:** 1- Rainger et al. (1990), 2- Milone et al (1991), 3- Christopoulou et al. (2012), 4- This Paper

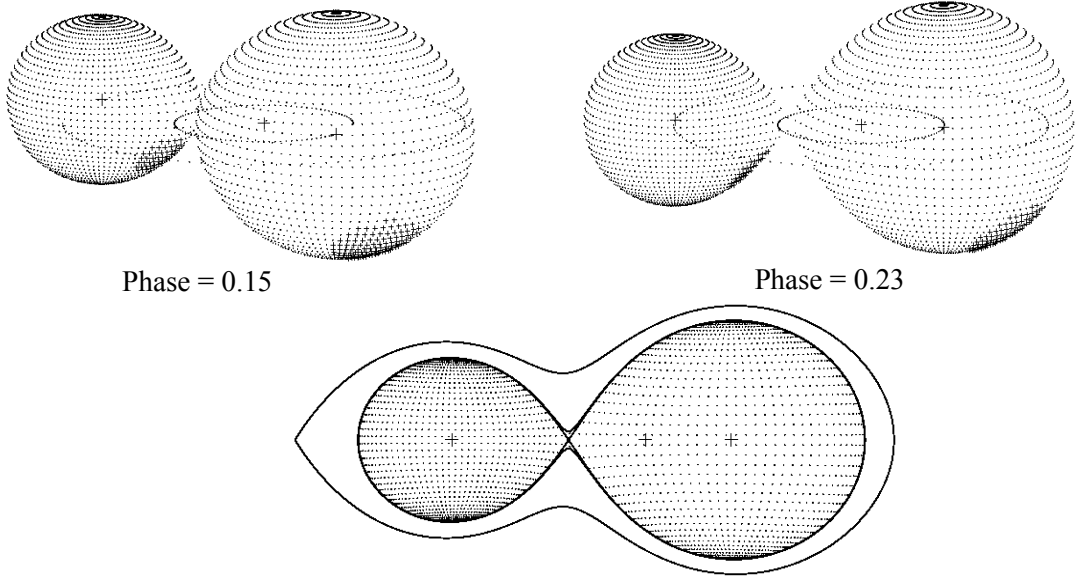

Figure 6: Geometric structure of the binary system TY Boo.

## 6. Evolutionary Status

In order to investigate the current evolutionary status of the system, we used the physical parameters listed in Table 7. We used the evolutionary tracks computed by Girardi et al. (2000) for both zero age main sequence stars (ZAMS) and terminal age main sequence stars (TAMS) with metalicity $z = 0.019$. The components of TY Boo are plotted on mass–luminosity (M-L) and mass–radius (M-R) relations, Figures 7 and 8. As it is clear from the Figures, the primary component of the system is located above the TAMS for the M-L relation while it located on the TAMS for M-R relation. The secondary component is close to the ZAMS track for both M-L and M-R relations. In the Figures, we plotted also the physical parameters computed by other authors listed in Table 7. The locations of both primaries and secondaries components are more or less have the same behavior as our solution. The same trend is obtained by Christopoulou et al. (2012) for a sample of W-type systems.

Location of our physical parameters on the mass-effective temperature relation ($M$–$T_{eff}$) relation for intermediate and low mass stars based on data of detached double-lined eclipsing binaries (Malkov, 2007) is displayed in Figure 9. The location of our mass and radius on the diagram revealed good fit for the secondary component and poor fit for primary one.

Some open questions are arisen when dealing with the evolutionary status of TY Boo system. Of them are the mass transfer from the secondary to the primary and its relation to the period change of the system. As the system has a low degree of contact, many authors (i.e. Christopoulou et al. (2012)) do not consider the mass transfer as a possible reason for period change.

Another open question is the influence of the magnetic activity on the period change. Our investigation shows that, the period change may be attributed to the magnetic activity. However, as shown by Stępień et al. (2001) the X-ray flux (for the time interval 1990-1991) weaker than that of single stars, so Christopoulou et al. (2012) excluded the magnetic activity as an occasion of period change.

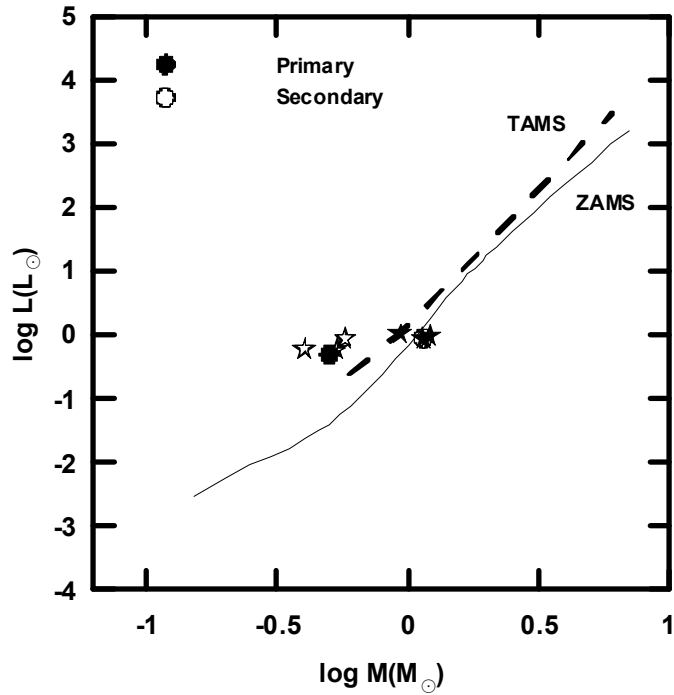

Figure 7: The position of the components of TY Boo on the mass–luminosity diagram. The filled circle denotes the primary and the open circle represents the secondary. Filled and open star symbols represent the masses and luminosities of the other solutions presented in Table 7.

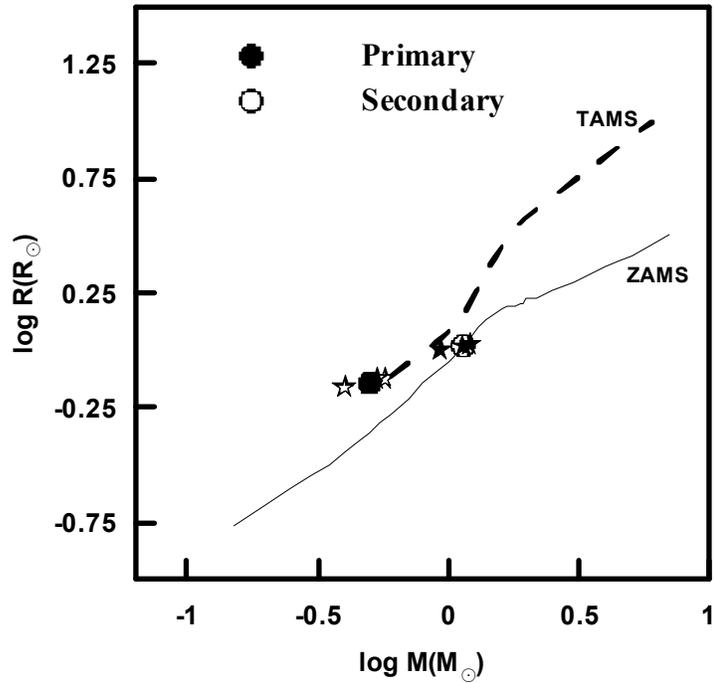

Figure 8: The position of the components of TY Boo on the mass–radius diagram. The filled circle denotes the primary and the open circle represents the secondary. Filled and open star symbols represent the masses and radii of the other solutions presented in Table 7.

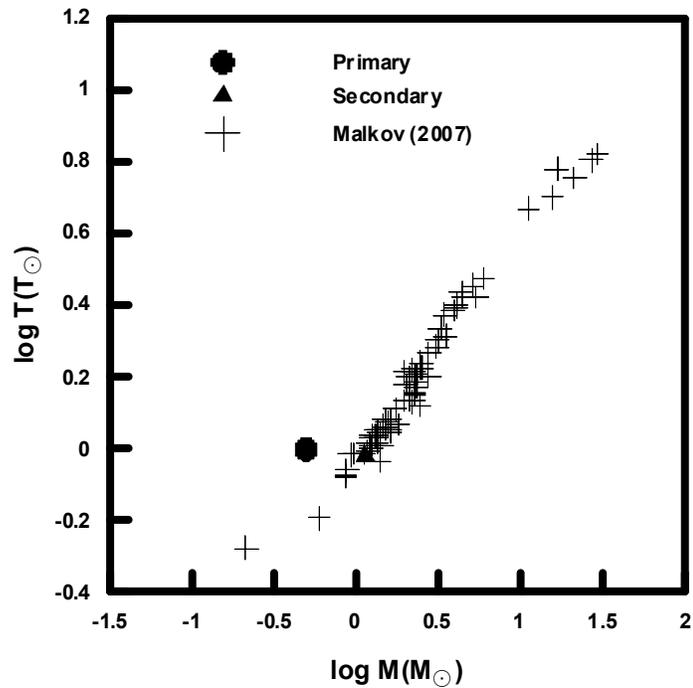

Figure 9. Position of the components of TY Boo on the empirical mass-Teff relation for low-intermediate mass stars by Malkov (2007).

7. **Discussion and Conclusion**

New CCD three colors light curves of TY Boo were carried out in five nights in BVR bandpass and new 19 times of minima were calculated. The parameters calculated by a photometric solution of these light curves by means of Willson-Devinney code showed that the low massive component is hotter than the more massive and cool one. The temperature difference between the components is $\Delta T \sim 249$ K. The period behavior of the system based shows a periodic behavior and a new light elements yields a new period (p= $0.^d3171506$) and shows a period decrease with the rate $dP/dE = 5.858 \times 10^{-12}$ day/cycle or $6.742 \times 10^{-9}$ day/year or 0.1 second/century. The conclusion reached may be drawn through the following points.

- Light curve modeling is performed using complete light curve in BVR bandpass. The curves were analyzed by means of W-D code and the accepted solution confirms the presence of hot spot on the hotter component and a dark spot on the cooler one. The absolute physical parameters were calculated and compared with that estimated by the previous light curve solutions for the system. The comparison showed that the physical properties of the system transform rapidly, which can interpreted by the Chromo spherical activity and rapid developing Ca II flare recorded in the system spectrum (Milone et al. 1990).
- We performed the first study of the long term stability of the system's TY Boo light curves and the possible connection with its period behavior, depending on all published light curves. Long term stability of the system light curves shows a periodic variation in the magnitude differences $D_{max}$, $D_{min}$, and the amplitude (depths) of the primary $A_p$ and secondary $A_s$.
- Synchronous of periodic variation for both orbital period and light curves parameters ($D_{max}$, $D_{min}$, $A_p$, and $A_s$) for the system TY Boo may be correlated by the presence of magnetic activity cycles and/or mass transfer mechanism. Future observations, particularly high resolution spectroscopic and photometric measurements are needed to follow this behavior.
- We have investigated the evolutionary status of the system using stellar models. The primary component is near or on the TAMS track, while the secondary component is still on the ZAMS track.